# Gate-tunable exciton-polaron Rydberg series with strong roton effect

Erfu Liu[1†], Jeremiah van Baren[1†], Zhengguang Lu[2,3], Takashi Taniguchi[4], Kenji Watanabe[5], Dmitry Smirnov[2], Yia-Chung Chang[6*], Chun Hung Lui[1*]

[1] Department of Physics and Astronomy, University of California, Riverside, California 92521, USA

[2] National High Magnetic Field Laboratory, Tallahassee, Florida 32310, USA

[3] Department of Physics, Florida State University, Tallahassee, Florida 32310, USA

[4] International Center for Materials Nanoarchitectonics, National Institute for Materials Science, 1-1 Namiki Tsukuba, Ibaraki 305-0044, Japan.

[5] National Institute for Materials Science, 1-1 Namiki Tsukuba, Ibaraki 305-0044, Japan.

[6] Research Center for Applied Sciences, Academia Sinica, Taipei 11529, Taiwan

[†] Equal contribution

[*] Corresponding authors. Emails: joshua.lui@ucr.edu; yiachang@gate.sinica.edu.tw

**The electronic exciton polaron is a hypothetical many-body quasiparticle formed by an exciton dressed with a polarized electron-hole cloud in the Fermi sea (FS). It is predicted to display rich many-body physics and unusual roton-like dispersion. Exciton polarons were recently evoked to explain the excitonic spectra of doped monolayer transition metal dichalcogenides (TMDs), but these studies are limited to the ground state. Excited-state exciton polarons can exhibit richer many-body physics due to their larger spatial extent, but detection is challenging due to their inherently weak signals. Here we observe gate-tunable exciton polarons for the 1s – 3s excitonic Rydberg series in ultraclean monolayer $MoSe_2$ devices by optical spectroscopy. When the FS expands, we observe increasingly severe suppression and steep energy shift from low to high Rydberg states. Their gate-dependent energy shifts go beyond the trion description but match our exciton-polaron theory. Notably, the exciton-polaron absorption and emission bands are separated with an energy gap, which increases from ground to excited state. Such peculiar characteristics are attributed to the roton-like exciton-polaron dispersion, where energy minima occur at finite momenta. The roton effect increases from ground to excited state. Such exciton-polaron Rydberg series with progressively significant many-body and roton effect shall provide a new platform to explore complex many-body phenomena.**

Optical spectra of excitons in the presence of Fermi sea (FS) have been a subject of fundamental interest for decades[1-3]. In a conventional scenario, an exciton captures an extra charge to form a three-body bound state called a trion, in analogy to the hydrogen ion.

The trion picture has been widely applied to explain the optical spectra of doped semiconductors, such as quantum wells[4-6], carbon nanotubes[7] and two-dimensional (2D) transition metal dichalcogenides (TMDs)[8-11]. Recent research, however, points out that the three-particle picture may be inadequate to account for the complex exciton-FS interactions[12-20]. In a more realistic description, an exciton can excite many electron-hole pairs in the FS. The exciton, dressed with the FS electron-hole pairs, can form a complex quasiparticle called exciton polaron[12-15] (not the conventional polarons dressed by phonons). The exciton polarons, as charge-neutral bosons, are fundamentally different from the trions, which are fermions with net electric charge.

There is on-going effort to distinguish the trion and exciton-polaron pictures, because they can produce similar optical spectra in the low-doping or weak-coupling regime. An exciton polaron can dissociate into a trion and a free hole when the coupling is weak. In the limit when the FS has only one electron, the exciton polaron is reduced to a trion. However, their behavior is distinct under strong exciton-FS coupling, which can be realized in two ways: (1) enlarging the FS (*e.g.* by electrostatic gating); (2) increasing the exciton size (*e.g.* by using excited-state excitons). Despite decades of exciton research, experimental investigation of *excited* excitonic states coupled to a *tunable* FS has been lacking thus far.

In this Letter, we apply both approaches to investigate the 1s – 3s excitonic states coupled to a gate-tunable FS in ultraclean monolayer $MoSe_2$. By using reflectance contrast and photoluminescence (PL) spectroscopy, we observe optical signatures of exciton-polaron Rydberg series and characterize their gate-dependent optical properties. We also established a comprehensive exciton-polaron theory to quantitatively explain our results. Our experiment and theory strongly support the exciton-polaron picture rather than the trion picture, especially in the regime of high charge density. In addition, the observed exciton polarons exhibit a significant energy gap between the absorption and emission bands, with the gap size increasing from ground to excited state. Such an absorption-emission gap can be attributed to the roton-like exciton-polaron dispersion, which becomes more pronounced in the excited state than the ground state. Our research reveals the crucial role of exciton polarons in doped 2D semiconductors and their increasingly complex behavior from ground to excited state.

We investigate ultraclean monolayer $MoSe_2$ gating devices encapsulated by hexagonal boron nitride (BN) with thin graphite as the contact and back-gate electrodes (see Methods and Fig. S1). Monolayer $MoSe_2$ possesses direct band gaps in two valleys, where the conduction (valence) band is split into two subbands with ~30 meV (~180 meV) separation by spin-orbit coupling[21]. The inner and outer subbands produce the A and B bright excitons, respectively (Fig. 1a). Each exciton hosts a series of internal energy levels, analogous to the hydrogenic Rydberg series (Fig. 1b)[22, 23]. We have measured the PL at magnetic fields $B = -31$ T to $+31$ T and gate voltage $V_g = 0$ V (Fig. 1c). We observe the A and B 1s excitons ($A_{1s}$, $B_{1s}$) and, between them, the A-exciton 2s and 3s states ($A_{2s}$, $A_{3s}$)[24-26]. From the magnetic-field-dependent exciton energies, we extract the linear Zeeman shift and quadratic diamagnetic shift[26-28]. $A_{2s}$ and $A_{3s}$ excitons exhibit noticeable diamagnetic shift, from which we can derive their root-mean-square radii $r_{2s} = 3.2$ nm and $r_{3s} = 8.1$ nm (Fig. 1d). The $A_{1s}$ diamagnetic shift is not observable in our experiment. We applied a model calculation to deduce the $A_{1s}$ Bohr radius $r_{1s} = 1.1$ $nm$, which matches the experimental value in the literature[26] (Table S1).

We have measured the gate-dependent reflectance contrast ($\delta R$) and PL maps, which reveal the absorption and emission properties of MoSe$_2$, respectively (Fig. 2, S2, S3, S4; see Methods). We take the second energy derivative of reflectance contrast $d^2(\delta R)/dE^2$ to enhance the weak features (Fig. 2e-h). We observe that the $A_{1s}$, $A_{2s}$, $A_{3s}$, $B_{1s}$ excitons subside on the electron (−) and hole (+) sides, and below them emerge new pairs of features labeled as $A_1^\pm$, $A_2^\pm$, $A_3^\pm$, $B_1^\pm$ respectively (Fig. 2; $A_3^\pm$ are observed only in PL). $A_1^\pm$ are commonly known as the trion states of $A_{1s}$ exciton. But $A_2^\pm$, $A_3^\pm$, and $B_1^\pm$ are new features observed in MoSe$_2$. A recent paper reported a trion excited state ($A_2^+$ or $A_2^-$) in monolayer WS$_2$, but the observation is limited to absorption with unknown doping level[29].

$A_1^\pm$, $A_2^\pm$, $A_3^\pm$, $B_1^\pm$ exhibit intriguing gate-dependent optical intensity. We first compare the $A_1^\pm$ and $B_1^\pm$ absorption intensity shown in the differential reflectance contrast maps (Fig. 2e-h). $A_1^\pm$ are suppressed at $|V_g| > 4$ V. Such gate-induced suppression is widely known, but the underlying mechanism is still uncertain. Here we can gain some insight by comparison with $B_1^\pm$. $B_1^\pm$ are less suppressed than $A_1^\pm$ and remain prominent on the hole side (Fig. 2f). Exciton suppression usually comes from two effects: (1) the plasma screening effect – free charges screen the Coulomb interaction; (2) the state-filling effect – carriers occupy the band-edge states that are needed to form excitons. A and B excitonic states experience similar screening effect as they share the same dielectric environment. However, they have different state-filling effect, because free carriers can easily block A excitons on the inner subbands, but not B excitons on the outer subbands (Fig. 1a, Fig. S5). This difference is more prominent on the hole side due to the large valence-band splitting. Correspondingly, we observe stronger gate-induced suppression on $A_1^\pm$ than $B_1^\pm$, on the hole side than electron side. The state-filling effect should therefore be the major suppression mechanism. Gate-induced suppression is also found in PL, but the A-B contrast is less pronounced (Fig. 2).

We next compare $A_1^\pm$, $A_2^\pm$ and $A_3^\pm$. Fig. 3a-b display the gate-dependent excitonic PL intensity. When the charge density increases, $A_{1s}$, $A_{2s}$, $A_{3s}$ quickly subside, but $A_1^\pm$, $A_2^\pm$, $A_3^\pm$ first grow and then diminish. Their respective critical charge density ($N_c$), defined at maximum PL, decreases monotonically from $N_1 \sim 3 \times 10^{12}$ cm$^{-2}$, $N_2 \sim 6.2 \times$ $10^{11}$ cm$^{-2}$ to $N_3 \sim 1.6 \times 10^{11}$ cm$^{-2}$ on the electron side (black triangles in Fig. 3a-b). The suppression increases from low to high Rydberg states. These observations can be roughly explained by the increasing state-filling effect from low to high states. High Rydberg excitons have larger spatial size and smaller *k*-space extent than low Rydberg excitons. From our extracted $A_{1s}$, $A_{2s}$, $A_{3s}$ exciton radii (inset of Fig.1d)[26], we estimate their *k*-space extent to be $k_{1s} \sim 0.44$ nm$^{-1}$, $k_{2s} \sim 0.16$ nm$^{-1}$ and $k_{3s} \sim 0.062$ nm$^{-1}$ by the uncertainty relationship $\Delta x \Delta k \sim 1/2$. For the conduction bands in monolayer MoSe$_2$ with valley degeneracy, the relation between carrier density and *k*-space extent is $N = k^2/2\pi$. The corresponding charge density to block these *k*-space decreases from $N_{1s} \sim 3.1 \times 10^{12}$ cm$^{-2}$, $N_{2s} \sim 3.9 \times 10^{11}$ cm$^{-2}$ to $N_{3s} \sim 6.1 \times 10^{10}$ cm$^{-2}$ (red dots in the inset of Fig. 3a). These values roughly match our measured critical charge density and hence support that $A_1^{\pm}$, $A_2^{\pm}$, $A_3^{\pm}$ are associated with the $A_{1s}$, $A_{2s}$, $A_{3s}$ excitons, respectively.

In addition to the gate-dependent suppression, the Rydberg states also exhibit remarkable gate-dependent energy shifts. Fig. 3c-d display the splitting energies $\Delta E_1$, $\Delta E_2$, $\Delta E_3$ between $A_1^{\pm}$, $A_2^{\pm}$, $A_3^{\pm}$ and $A_{1s}$, $A_{2s}$, $A_{3s}$, respectively, as a function of gate voltage $V_g$, charge density *N*, and Fermi level $E_F$. The splitting energies are different for PL (Fig. 3c) and reflection (Fig. 3d). For PL, $\Delta E_1$, $\Delta E_2$ and $\Delta E_3$ range from 15 to 40 meV in our measurements. All of them exhibit approximately linear increase with $E_F$; their slope with $E_F$ increases from $0.93 \pm 0.04$, $9.3 \pm 0.3$ to $11.5 \pm 1.0$, respectively (Fig. 3c). Linear extrapolations give $\Delta E_1 = 26.1$ meV and $\Delta E_2 = 24.6$ meV, $\Delta E_3 = 13.0$ meV at the charge neutrality. For reflection, $\Delta E_1$ and $\Delta E_2$ range from 20 to 35 meV and both increase with $E_F$. Their slope with $E_F$ increases from $2.4 \pm 0.1$ for $\Delta E_1$ to $5.9 \pm 0.5$ for $\Delta E_2$. Linear extrapolations give $\Delta E_1 = 23.1$ meV and $\Delta E_2 = 16.4$ meV at the charge neutrality. The increasing slope of $\Delta E_1$, $\Delta E_2$, $\Delta E_3$ indicates progressively increasing exciton-FS interaction from low to high Rydberg states.

To clarify the nature of $A_1^{\pm}$, $A_2^{\pm}$, $A_3^{\pm}$ states, we have calculated the absorption resonances of the $A_{1s}$, $A_{2s}$, $A_1^-$, $A_2^-$ states by first-principles calculations based on both the trion picture and exciton-polaron picture (see details in the Supplementary Information). Our calculations include the screening and state-filling effect as well as the band-structure renormalization by the FS. The calculated $A_1^-$ trion binding energy decreases with $V_g$ due to the increasing charge screening and state-filling effect at higher carrier density (dashed lines in Fig. 4d). The result contradicts our observation that both $\Delta E_1$ and $\Delta E_2$ increase with carrier density. The trion picture is therefore not applicable to explain our data.

Our calculations based on the exciton-polaron theory produce significantly different results (Fig. S8). In our calculation, we only consider the dominate component of an exciton-polaron – the Suris tetron, which is an exciton bound with one FS electron-hole pair[30, 31] (Fig. 4e). Such simplification is justified because, after an exciton becomes a Suris tetron, its coupling to the FS is much reduced. A Suris tetron can be considered as a trion bound with a FS hole (in the case of electron FS). When the electron FS expands, the enlarged phase space of the FS hole can increase the trion-hole binding so that the Suris tetron becomes more strongly bound at higher carrier density. Fig. 3d displays the calculated splitting energy between $A_{1s}$, $A_{2s}$ excitons and their associated Suris tetrons. Our calculations quantitatively reproduce the increase of splitting energy with charge density for both the ground and excited states. The agreement with experiment strongly supports the exciton-polaron picture.

We next consider the marked difference between the PL and absorption energies in the exciton-polaron states (Fig. S6). Fig. 4 displays the gate-dependent energy difference $\Delta E_{R-PL}$ between the reflection and emission bands for $A_1^\pm$ and $A_2^\pm$. For $A_1^\pm$, $\Delta E_{R-PL}$ is nearly zero at $|V_g| < 3$ V, but increases almost linearly at $|V_g| > 3$ V. For $A_2^\pm$, $\Delta E_{R-PL}$ increases linearly from $|V_g| \sim 0$ V, with a steeper slope than that of $A_1^\pm$.

The large absorption-emission energy gap can be attributed to the roton-like exciton-polaron dispersion, which has been predicted by recent theory[20]. In the absence of FS, the excitons have quadratic dispersion; their absorption and emission have similar energy because both mainly occur at the energy minimum with zero center-of-mass momentum (inset of Fig. 4). In the presence of FS, however, the Pauli blocking effect will restrict the optical absorption to generate electron-hole pairs away from the band edge; such excitons still have zero momentum. But the hole can relax to the band edge to lower the exciton energy; such excitons have finite center-of-mass momentum. The interplay of the hole relaxation and many-body interactions can shift the exciton energy minimum from zero to finite momentum (inset of Fig. 4). Therefore, the exciton-FS coupling can produce a roton-like dispersion. While exciton absorption still occurs at zero momentum, the emission occurs predominantly at the roton minimum with finite momentum. As momentum-indirect emission requires scattering with defects, phonons or ambient carriers, the emission intensity is expected to diminish. Therefore, the roton effect has two signatures: (1) the absorption energy is larger than the emission energy; (2) the PL intensity diminishes. We observe both signatures in our experiment (Fig. 3b, 4). Notably, the excited-state polarons ($A_2^\pm$) have considerably stronger roton effect than ground-state polarons ($A_1^\pm$), because the exciton-FS interaction is stronger in the excited state than ground state.

Finally, we comment on the $A_3^\pm$ states. As we only observe $A_3^\pm$ in the PL, we cannot directly determine its absorption-emission gap $\Delta E_{R-PL}$. But compared to $A_1^\pm$ and $A_2^\pm$, $A_3^\pm$ exhibit the largest PL redshift slope and strongest suppression with increasing $E_F$, which indicate the strongest exciton-FS interaction. We therefore speculate that $A_3^\pm$ also have the strongest roton effect among the three. Further research is merited to investigate the many-body and roton effects in the high-lying excited exciton-polaron states.

In summary, we have observed the Rydberg series of A excitonic states ($A_1^\pm$, $A_2^\pm$, $A_3^\pm$) at gate-tunable electron and hole density. Our analysis supports that these states are exciton polarons rather than trions. From the ground to excited states, the exciton polarons exhibit increasingly strong gate-induced suppression and absorption-emission energy gap. The results reveal that the interaction with Fermi sea, the state-filling effect and the roton effect, become increasingly strong from the ground to excited states of the exciton-polaron. Besides, we have also observed ground and excited exciton polarons in monolayer $WSe_2$ (Fig. S7) with similar characteristics as those in monolayer $MoSe_2$; the results strongly suggest that exciton polarons generally exist in many 2D semiconductors. Our observation of exciton-polaron Rydberg series and their many unusual properties shall open a new path to exploring novel many-body physics in 2D systems.

Note added. – During the preparation of our manuscript, two related papers were submitted[32, 33].

**Acknowledgments:** We thank H. Yu, W. Yao, A. P. Mills, A. H. MacDonald and D. K. Efimkin for discussions, N. M. Gabor and Y. Cui for co-supporting E.L, and H. W. K. Tom for equipment support. C.H.L. acknowledges support from the National Science Foundation Division of Materials Research CAREER Award No.1945660. Y.C.C. is supported by Ministry of Science and Technology (Taiwan) under grant Nos. MOST 107-2112-M-001-032 and 108-2112-M-001-041. K.W. and T.T. acknowledge support from the Elemental Strategy Initiative conducted by the MEXT, Japan and the CREST (JPMJCR15F3), JST. A portion of this work was performed at the National High Magnetic Field Laboratory, which is supported by the National Science Foundation Cooperative Agreement No. DMR-1644779 and the State of Florida.

**Author contributions:** E.L. fabricated the devices. J.v.B established experimental facilities for the research. E.L. and J.v.B. performed the experiments and analyzed the data. Z.L. and D.S. supported the magneto-optical experiment. T.T. and K.W. provided boron nitride crystals for device fabrication. Y.C.C. did the theoretical calculations. C.H.L. supervised the project. C.H.L., E.L. and Y.C.C. wrote the manuscript.

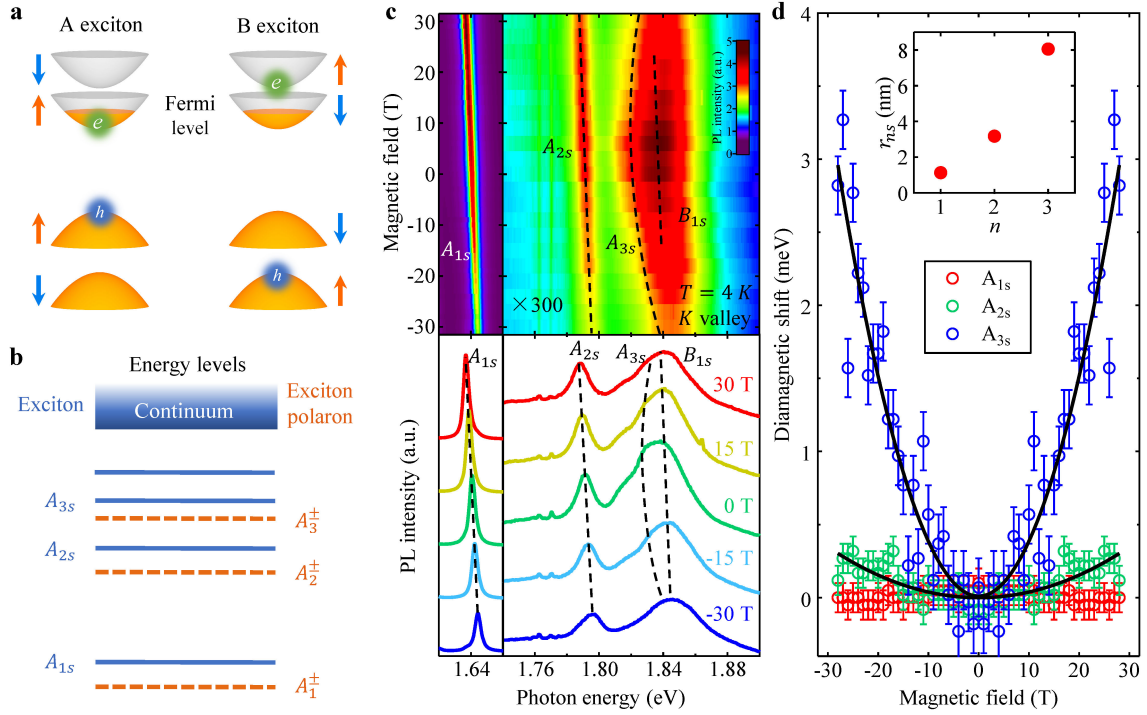

**Figure 1 | Excitonic states in monolayer MoSe$_2$.** **a.** The band configurations of A and B bright excitons in monolayer MoSe$_2$. The arrows denote the electron spin of the bands. The Fermi level reaches only the inner subbands in our experiment. **b.** The internal energy levels of exciton and exciton polaron. **c.** Magnetic-field-dependent photoluminescence (PL) maps of monolayer MoSe$_2$ at zero gate voltage and temperature $T$ = 4 K. The lower panel shows the cross-cut PL spectra at five different magnetic fields. The dash lines highlight the exciton energy shifts. The spectra in 1.71 – 1.92 eV are magnified by 300 times for clarity. **d.** The diamagnetic energy shift of the A-exciton 1s, 2s and 3s states as a function of magnetic field. The lines are quadratic fits. The inset shows the root-mean-square exciton radii ($r_{ns}$) for principal quantum number $n$ = 1, 2, 3. $r_{2s}$ and $r_{3s}$ are extracted from the diamagnetic shifts. $r_{1s}$ is obtained by a theoretical model and it matches the experimental value in Ref. 26.

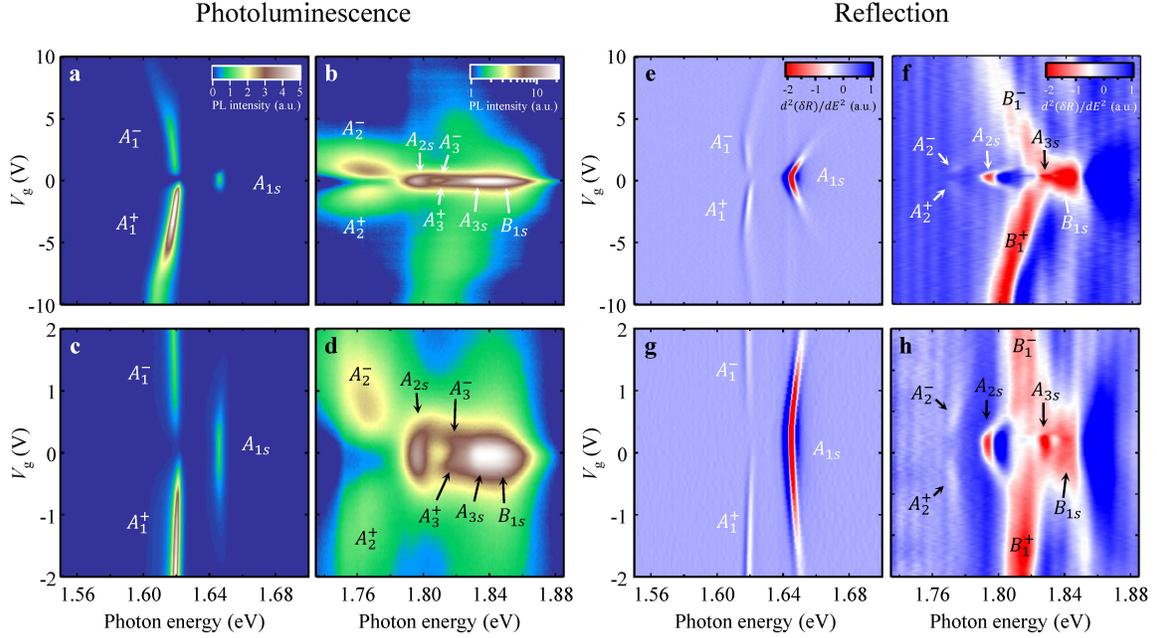

**Figure 2 | Observation of exciton-polaron Rydberg states in monolayer MoSe$_2$. a, b,** Gate-dependent photoluminescence (PL) maps. **c, d,** Zoom-in PL maps at gate voltage $V_g$ = -2 to +2 V. **a** and **c** share the same color scale bar; **b** and **d** share the same log-scale color bar. **e, f,** Gate-dependent maps of the second energy derivative of reflectance contrast $d^2(\delta R)/dE^2$. **g, h,** Zoom-in $d^2(\delta R)/dE^2$ maps at $V_g$ = -2 to +2 V. **e-h** share the same color scale bar. The spectra in panels **b**, **d** (**f**, **h**) are magnified for 300 (20) times, respectively. The measurements were made at sample temperature $T \sim 15$ K with no magnetic field.

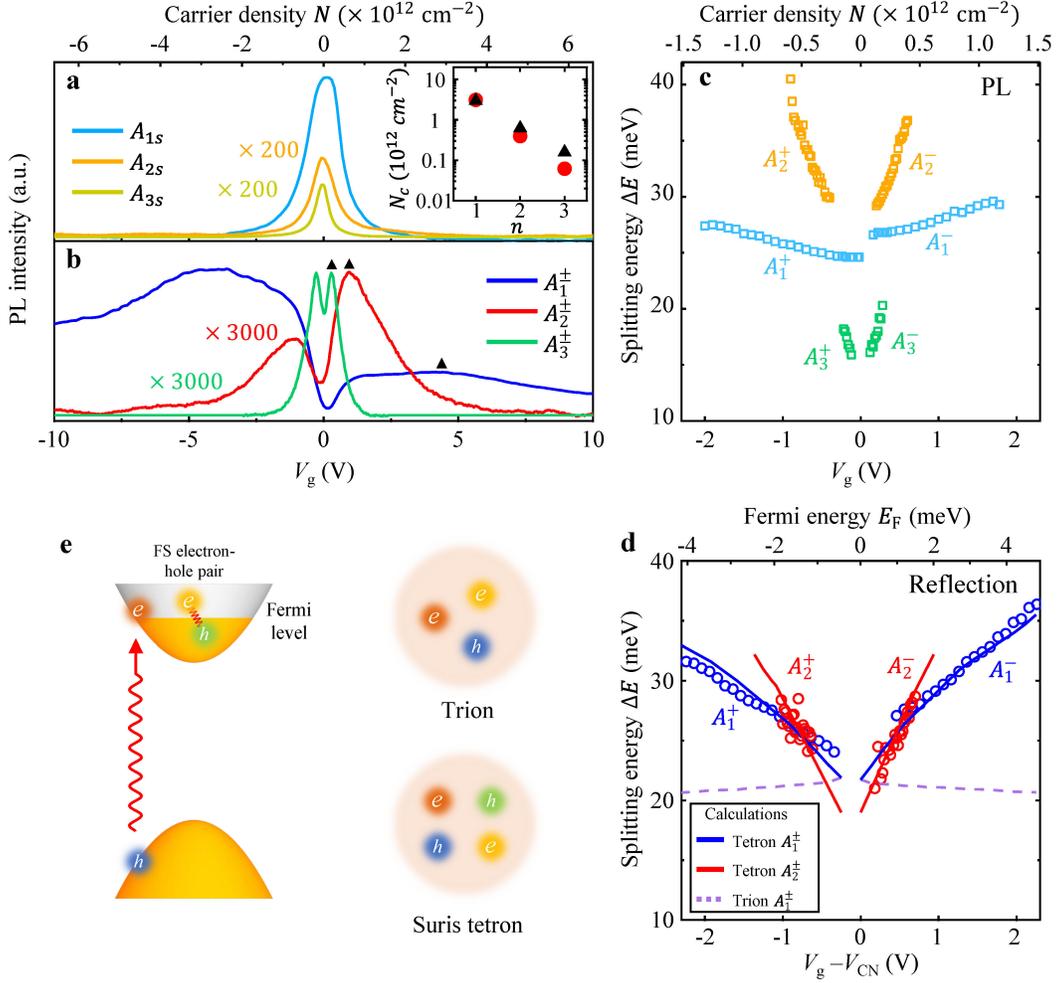

**Figure 3 | Gate-dependent optical characteristics of ground and excited exciton polarons in monolayer MoSe$_2$. a, b,** Gate-dependent integrated PL intensity of excitons and exciton polarons as a function of gate voltage (bottom axis) and charge density (top axis). The charge neutrality point is at $V_g = 0$ V. We magnify the $A_{2s}$, $A_{3s}$ ($A_2^{\pm}$, $A_3^{\pm}$) PL intensity for 200 (3000) times for clarify. The inset shows the charge density at maximum PL intensity in panel **b** (black triangles), compared to the quenching density (red dots) expected from the state-filling effect with the exciton radii in Fig. 1d. **c,** Gate-dependent PL splitting energy between $A_1^{\pm}$, $A_2^{\pm}$, $A_3^{\pm}$ and $A_{1s}$, $A_{2s}$, $A_{3s}$, respectively. **d,** Gate-dependent absorption splitting energy between $A_1^{\pm}$, $A_2^{\pm}$ and $A_{1s}$, $A_{2s}$ reflectance contrast features, respectively. The charge neutrality point is at $V_{CN} \sim 0.25$ V, which has been subtracted from $V_g$. The dashed lines are the calculated trion binding energy. The solid lines are the fitting by our theoretical model of exciton polarons. **e,** Schematic of exciton-FS interaction that leads to the exciton-polaron formation. A photo-excited exciton can excite electron-hole pairs on the Fermi sea (FS). The exciton, coupled with one FS electron-hole pair, form a Suris tetron, in contrast to a trion with only three particles.

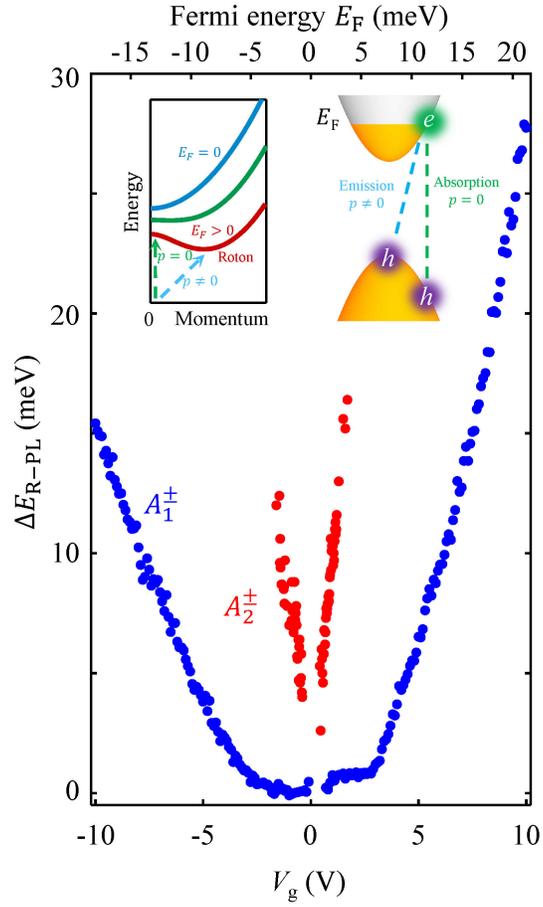

**Figure 4 | Signature of roton effect in the exciton polarons.** Energy separation between the reflection and emission features as a function of gate voltage for both the $A_1^\pm$ and $A_2^\pm$ exciton polarons. Inset illustrates the roton-like exciton-polaron dispersion, where the energy minima occur at finite momenta due to the state blocking of the Fermi sea. Such dispersion leads to different energies for optical absorption and emission.

**Methods**

**Device fabrication.** We fabricate monolayer MoSe$_2$ devices encapsulated by hexagonal boron nitride (BN) by stacking different component 2D materials together. We first exfoliate monolayers MoSe$_2$, multilayer graphene and thin BN flakes from their bulk crystals onto the Si/SiO$_2$ substrates (The MoSe$_2$ crystals are bought from HQ Graphene Inc.). Afterward, we apply the polycarbonate-based dry-transfer technique [34, 35] to stack them together. In this method, we use a stamp to first pick up a BN flake, and then use the BN flake to sequentially pick up two pieces of multilayer graphene (as the contact electrodes), a MoSe$_2$ monolayer, a BN thin layer (as the bottom gate dielectric), and a graphene multilayer (as the back-gate electrode). Our method ensures that the MoSe$_2$ layer doesn't contact the polymers during the whole fabrication process, so as to reduce the contaminants and bubbles at the interfaces[36]. Standard electron beam lithography is then applied to pattern and deposit the gold contacts (100-nm thickness). Finally, we anneal the devices at 300 °C for three hours in an argon environment. Fig. S1 shows the schematic and microscopy image of a representative BN-encapsulated monolayer MoSe$_2$ device.

**Experimental methods.** The optical experiments with no magnetic field (Fig. 2-4) were performed in our laboratory at UC Riverside. We mount the devices in a cryostat (Montana Instrument) with sample temperature at $T \sim 15$ T. For the reflectance contrast measurements, we focus the white light from a broadband light source (Thorlabs, SLS201L) onto the sample with a spot diameter of ~2 μm. We measure the reflected spectrum ($R_s$) from the monolayer MoSe$_2$ sample on the BN/MoSe$_2$/BN/Gr/SiO$_2$/Si stack, and the reference spectrum ($R_r$) on a nearby area without MoSe$_2$ on the BN/BN/Gr/SiO$_2$/Si stack. The reflectance contrast ($\delta R$) is obtained as $\delta R = (R_s - R_r)/R_r$. We further take the second energy derivative of the $\delta R$ spectrum to sharpen the weak features.

For the photoluminescence (PL) experiment with no magnetic field, we excite the samples with a 532-nm continuous laser (Laser Quantum; Torus 532). The laser is focused through a microscope objective (NA = 0.6) onto the sample with a spot diameter of ~1 μm. The PL is collected through the same objective in a backscattering geometry and analyzed by a high-resolution spectrometer with a charge-coupled device (CCD) camera (Princeton Instruments).

The PL experiments with magnetic field were performed in the National High Magnetic Field Laboratory (NHMFL) in Florida, USA. We use a 31-Tesla DC magnet and a fiber-based probe (the setup is the same as that in Ref. 27). The sample temperature is $T = 4$ K. A 532-nm continuous laser is directed through a single-mode optical fiber and focused by a lens (NA = 0.67) onto the sample. The sample is mounted on a three-dimensionalpiezoelectric translational stage (attocube). The PL is collected through a 50/50 beam splitter into a multimode optical fiber, and subsequently measured by a spectrometer (Princeton Instruments, IsoPlane 320) with a CCD camera.